\documentclass[aps,pre,twocolumn,showpacs,superscriptaddress,groupedaddress]{revtex4}  
\usepackage{graphicx}  
\usepackage{dcolumn}   
\usepackage{bm}        
\usepackage{amssymb}   
\usepackage{color}
\begin{document}
\widetext
\title{Frequency spectra of turbulent thermal convection with uniform rotation}
\author{Hirdesh K. Pharasi}
\author{Krishna Kumar\footnote{\it kumar@phy.iitkgp.ernet.in}}
\affiliation{Department of Physics, Indian Institute of Technology, Kharagpur-721 302, India}
\author{Jayanta K. Bhattacharjee}
\affiliation{Harish-Chandra Research Institute, Allahabad-211 019,  India} 
\date{\today}
\begin{abstract}
The frequency spectra of the  entropy and kinetic energy along with the power spectrum of the thermal flux are computed from direct numerical simulations for turbulent Rayleigh-B\'{e}nard convection with uniform rotation about a vertical axis in low-Prandtl-number fluids ($\mathrm{Pr} < 0.6$). Simulations are done for convective Rossby numbers $\mathrm{Ro} \ge 0.2$. The temporal fluctuations of these global quantities show two scaling regimes: (i) $\omega^{-2}$ at higher frequencies for all values of $\mathrm{Ro}$ and (ii) $\omega^{-\gamma_1}$ at intermediate frequencies with $\gamma_1 \approx 4$  for $\mathrm{Ro} > 1$, while $4 < \gamma_1 < 6.6$ for $0.2 \le \mathrm{Ro} < 1$. 
\end{abstract}
\maketitle
In the Kolmogorov picture (K41) of homogeneous isotropic turbulence, kinetic energy is dissipated at short length scales (large wave numbers $k$) by viscosity. For a statistically steady state to be obtained, the kinetic energy is injected at large length scales at a constant rate which is also the dissipation rate. Far away from either end is the inertial range through which the nonlinear terms of the Navier-Stokes equation transfer the energy from one scale to another at a constant rate $\epsilon_K = dK/dt$, where $K = \int{ E(k)d\mathbf{k}} = \int{d\mathbf{k} d\omega C_K (\mathbf{k}, \omega)}$ is the kinetic energy per unit mass and $C_K (\mathbf{k}, \omega)$ is the wave number $k$ and frequency $\omega$ dependent velocity correlation function. The object $\epsilon_K$ is thus the kinetic energy flux. In the inertial range, $E(k)$ is dependent only on wave number $k$ and $\epsilon_K$. A dimensional analysis leads to $E(k)\sim k^{-5/3}$~\cite{K41} (K41). In general, $E(k) = 4\pi k^2 \int{C_K (\mathbf{k}, \omega) d\omega} \propto k^{-\alpha}$. Scaling implies $C_K (\mathbf{k}, \omega) = k^{-2-\alpha-\zeta} f(\omega/k^{\zeta})$, where $k^{\zeta}$ is a characteristic frequency. A dimensional argument yields the exponent $\zeta = 2/3$ in K41.

In convective turbulence~\cite{turbulence_review, lohse_etal_review}, one expects that there will be a regime where temperature fluctuations $\delta T$, rather than velocity fluctuations $v$, will control the spectrum. It was shown by L'vov~\cite{vsl_1991} that for small temperature fluctuations, the expansion of the mean convective entropy $\Phi \propto \int (\delta T)^2 d \mathbf{r}$, where $\mathbf{r}$ is the space variable. The  entropy will be conserved in the absence of thermal diffusion and forcing. In the so called Bolgiano-Obukhov~\cite{BO} picture (BO) in the corresponding inertial range, $E(k)$ is determined by $k$ and the rate of dissipation of the entropy $\epsilon_\Phi = d\Phi/dt$ which is equal to the flux of the convective entropy through different scales. Here $E(k)$  is determined by $k$, $\epsilon_\Phi$ and $\alpha_0 g$, where $\alpha_0$ is the thermal expansion coefficient. One gets $E(k)\sim k^{-11/5}$, which gives $\alpha = 11/5$ in this case. The corresponding $\zeta$ is known to be $2/5$. There will be a convective entropy spectrum $E_\Phi(k)$ as well defined by $\Phi = \int E_\Phi(k) d\mathbf{k}$. Dimensional analysis establishes that the entropy spectrum $E_\Phi(k)\sim k^{-7/5}$. 

In any realistic situation K41 [$E(k)\sim k^{-5/3}$] and BO [$E(k)\sim k^{-11/5}$] will both be present with K41 holding at high wave numbers and BO at smaller ones. The two exponents are quite close to each other and attempts to go to small enough $k$ to get into the BO regime has been thwarted by the system size. Consequently, we decided to explore the spectra in frequency space instead. 

 The frequency spectra $E (\omega)$ for the kinetic energy and $E_{\Phi} (\omega)$ for the convective entropy are defined as: $K = \int{E(\omega) d\omega}$ and $\Phi = \int{E_{\Phi} (\omega) d\omega}$. In terms of correlation functions $E(\omega) = 4\pi \int{k^2 C_K (\mathbf{k}, \omega) d\mathbf{k}}$ and $E_{\Phi} (\omega) = 4\pi \int{k^2 C_{\Phi} (\mathbf{k}, \omega) d\mathbf{k}}$. Insertion of the scaling form for $C_K$ shows $E (\omega) \sim \omega^{-2}$ for K41 and $\omega^{-4}$ for BO. On the other hand the scaling form for $C_{\Phi}$ shows $E_{\Phi} (\omega) \sim \omega^{-2}$ for BO and $\omega^{-4}$ for K41. In reality $E (\omega)$ and $E_{\Phi} (\omega)$ get contributions from both K41 and BO, and to the lowest order we find that the two contributions add. The relative contributions of K41 and BO for $E$ and $E_{\Phi}$ are of a similar form, with BO a weaker contribution. We write $E (\omega) \approx \omega^{-2} + c_0 \omega_o^2 \omega^{-4}$, where $c_0 <1$ for a weak BO contribution, while $\omega_o$ depends on system parameters. For the convective entropy, $E_{\Phi} (\omega) \approx c_0 \omega^{-2} + \omega_o^2 \omega^{-4}$. For $E (\omega)$, the crossover frequency is $\omega_K = \omega_o \sqrt{c_0}$ and for $E_{\Phi} (\omega)$ it is $\omega_{\Phi} = \omega_o/\sqrt{c_0}$. Since $c_0 < 1$, the crossover should occur earlier for $E (\omega)$. The dependence of $c_0$ on the rotation rate cannot be predicted.

\begin{table*}[ht]
\caption{\label{table1} List of the convective Rossby number $\mathrm{Ro} =\sqrt{\mathrm{Ra/(Ta Pr)}}$, the Prandtl number $\mathrm{Pr}$, the reduced Rayleigh number $r = \mathrm{Ra/Ra_{\circ}(Ta, Pr)}$, the Rayleigh number $\mathrm{Ra}$, and the scaling exponents ${\gamma_1}$ at intermediate frequencies and ${\gamma_2}$ at higher frequencies for the power spectra of three global quantities: the entropy, the kinetic energy, and the Nusselt number.}
\begin{ruledtabular}
\begin{tabular}{cccccccccc}
$\mathrm{Ro}$ & $\mathrm{Pr}$ &  $r$ & $\mathrm{Ra}$& $\gamma_1^{\Phi}$ & $\gamma_2^{\Phi}$ & $\gamma_1$ & $\gamma_2$ & $\gamma_1^{\mathrm{Nu}}$ & $\gamma_2^{\mathrm{Nu}}$    \\ \hline
$0.20$ & $0.1$ & $5$ & $4.03 \times 10^6$  & $6.01 \pm 0.27$ & $1.97$ & $6.30 \pm 0.24$ & $1.97$& $5.86 \pm 0.27$ & $1.98$\\
$0.30$ & $0.1$ & $5$ & $8.97 \times 10^5$  & $6.33 \pm 0.26$ & $1.98$ & $6.20 \pm 0.17$ & $1.97$& $6.24 \pm 0.11$ & $1.97$\\
$0.42$ & $0.1$ & $10$ & $1.79 \times 10^6$  & $4.66 \pm 0.12$ & $1.97$ & $4.54 \pm 0.18$ & $1.98$& $4.63 \pm 0.12$ & $1.97$\\
$0.52$ & $0.5$ & $10$ & $1.33 \times 10^7$  & $4.31 \pm 0.14$ & $1.97$ & $4.36 \pm 0.11$ & $1.98$& $4.33 \pm 0.09$ & $1.97$\\
$0.73$ & $0.1$ & $30$ & $5.38 \times 10^6$  & $4.06 \pm 0.13$ & $1.97$ & $4.28 \pm 0.10$ & $1.97$& $3.87 \pm 0.10$ & $1.98$\\
$1.34$ & $0.1$ & $100$ & $1.79 \times 10^7$  & $3.67 \pm 0.12$ & $1.97$ & $3.96 \pm 0.19$ & $1.97$& $3.67 \pm 0.09$ & $1.97$  \\
$3.27$ & $0.1$ & $100$ & $1.07 \times 10^6$  & $4.12 \pm 0.09$ & $1.96$ & $4.25 \pm 0.13$ & $1.97$ & $3.83 \pm 0.06$ & $1.97$  \\
$7.70$ & $0.5$ & $100$ & $8.90 \times 10^5$  & $3.48 \pm 0.11$ & $1.97$ & $3.66 \pm 0.17$ & $1.96$ & $3.66 \pm 0.09$ & $1.96$\\
$10.26$ & $0.5$ & $100$ & $5.26 \times 10^5$  & $3.95 \pm 0.15$ & $1.97$ & $3.97 \pm 0.13$ & $1.97$ & $4.12 \pm 0.10$ & $1.96$ \\
$23.13$ & $0.1$ & $5000$ & $5.35 \times 10^7$  & $3.60 \pm 0.10$ & $1.97$ & $3.84 \pm 0.19$ & $1.97$ & $3.70 \pm 0.07$ & $1.97$  \\
$42.19$ & $0.5$ & $3000$ & $2.67 \times 10^7$  & $3.89 \pm 0.11$ & $1.97$ & $4.25 \pm 0.12$ & $1.97$ & $3.77 \pm 0.02$ & $1.97$\\
$56.21$ & $0.5$ & $3000$ & $1.58 \times 10^7$  & $4.01 \pm 0.11$ & $1.97$ & $3.38 \pm 0.15$ & $1.97$ & $3.62 \pm 0.05$ & $1.96$ \\
$72.66$ & $0.1$ & $3000$ & $5.28 \times 10^6$  & $4.08 \pm 0.16$ & $1.97$ & $3.26 \pm 0.20$ & $1.97$ & $3.82\pm 0.06$ & $1.96$ \\
$\infty$ & $0.1$ & $3000$ & $2.00 \times 10^6$ & $4.13 \pm 0.17$ & $1.97$ & $3.62 \pm 0.14$ & $1.97$ & $4.11 \pm 0.08$ & $1.97$  \\
\end{tabular}
\end{ruledtabular}
\end{table*}

\begin{figure*}[ht]
\includegraphics[height=10 cm, width=15 cm]{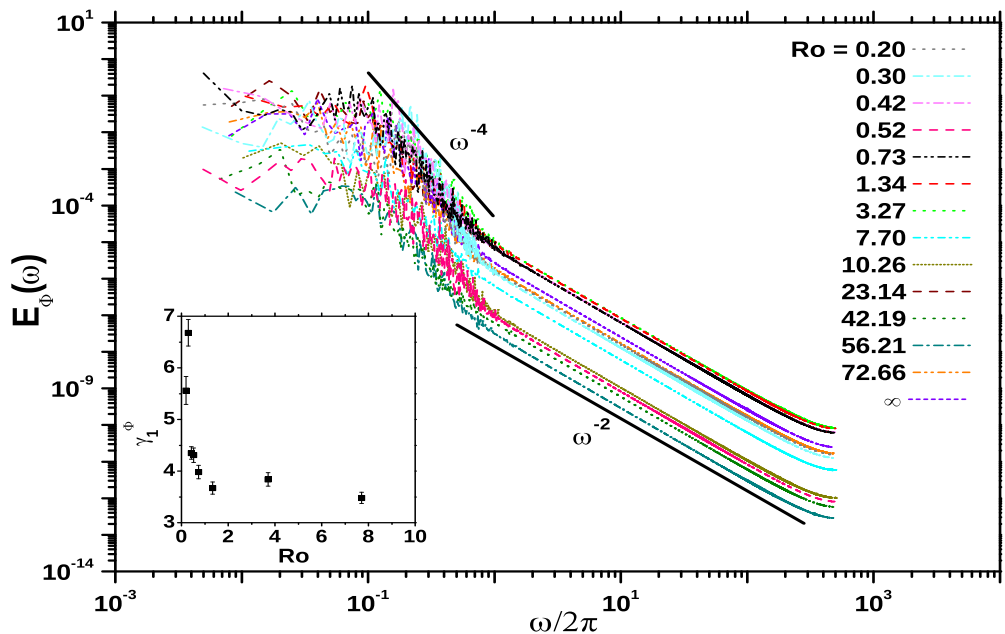}
\caption{(Color online) Frequency spectra $E_{\Phi} (\omega)$ of the spatially averaged convective entropy computed from DNS for different values of $\mathrm{Ro}$ (curves of different colors). The spectra show two scaling regimes: (i) $E_{\Phi} (\omega)$ scales as $\sim \omega^{-\gamma_1^{\Phi}}$ at the moderate frequencies, with the exponent $\gamma_1^{\Phi}$ showing a weak dependence on $\mathrm{Ro}$ for $\mathrm{Ro} > 1$ ($3.37 \le \gamma_1 \le 4.30$), and a strong dependence on $\mathrm{Ro}$ for $\mathrm{Ro} < 1$ ($3.93 \le \gamma_1 \le 6.59$). (ii) $E_{\Phi} (\omega)$ scales with $\omega$ as $\sim \omega^{-\gamma_2^{\Phi}}$ with $\gamma_2^{\Phi} \sim 2$ at higher frequencies for all values of $\mathrm{Ta}$, $\mathrm{Pr}$, and $\mathrm{Ro}$ considered here.  The upper and lower black straight lines show the scaling behavior as $\omega^{-4}$ and $\omega^{-2}$, respectively. The inset shows the variation of the exponent $\gamma_1^{\Phi}$ with $\mathrm{Ro}$.} \label{entropy} 
\end{figure*}

\begin{figure*}[ht]
\includegraphics[height=10 cm, width=15 cm]{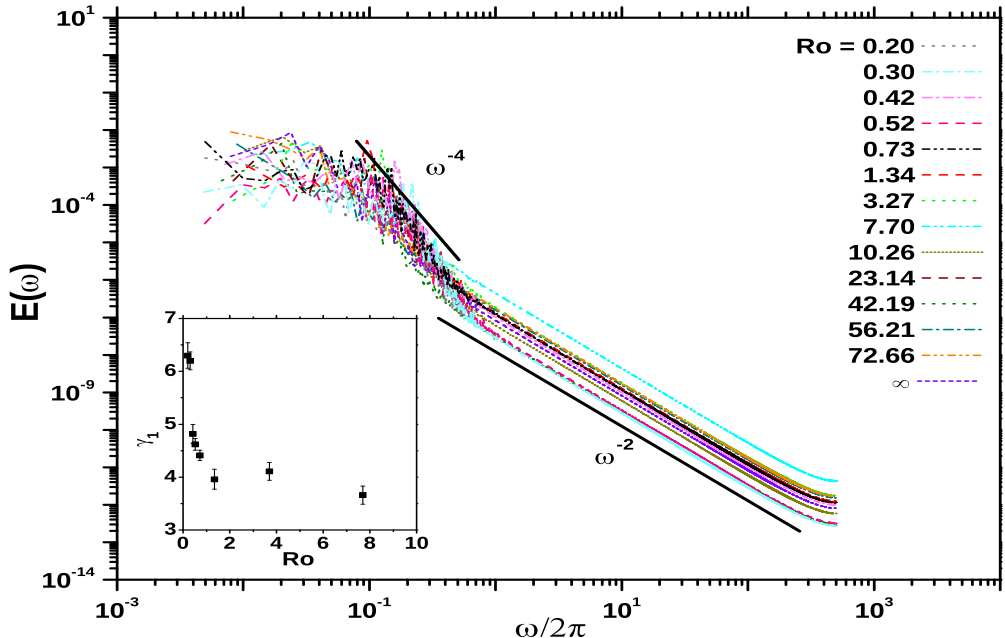}
\caption{(Color online) Frequency spectra $E (\omega)$ of the spatially averaged kinetic energy, as computed from DNS, for the several values of the Rossby number $0.2 \le \mathrm{Ro} < \infty$: $E(\omega)$ scales with $\omega$ as $\omega^{-\gamma_1}$ at moderate frequencies with $3.06 \le \gamma_1 \le 6.54$ and $E (\omega)$ scales with $\omega$ as $\omega^{-\gamma_2}$ with $\gamma_2 \sim 2$. The exponent $\gamma_2$ is found to be independent of $\mathrm{Ta}$, $\mathrm{Pr}$, and $\mathrm{Ro}$. The upper and lower black lines stand for the scaling behaviors $\omega^{-4}$ and $\omega^{-2}$, respectively. The inset displays the variation of $\gamma_1$ with $\mathrm{Ro}$.} \label{energy} 
\end{figure*}

\begin{figure*}[ht]
\includegraphics[height=10 cm, width=15 cm]{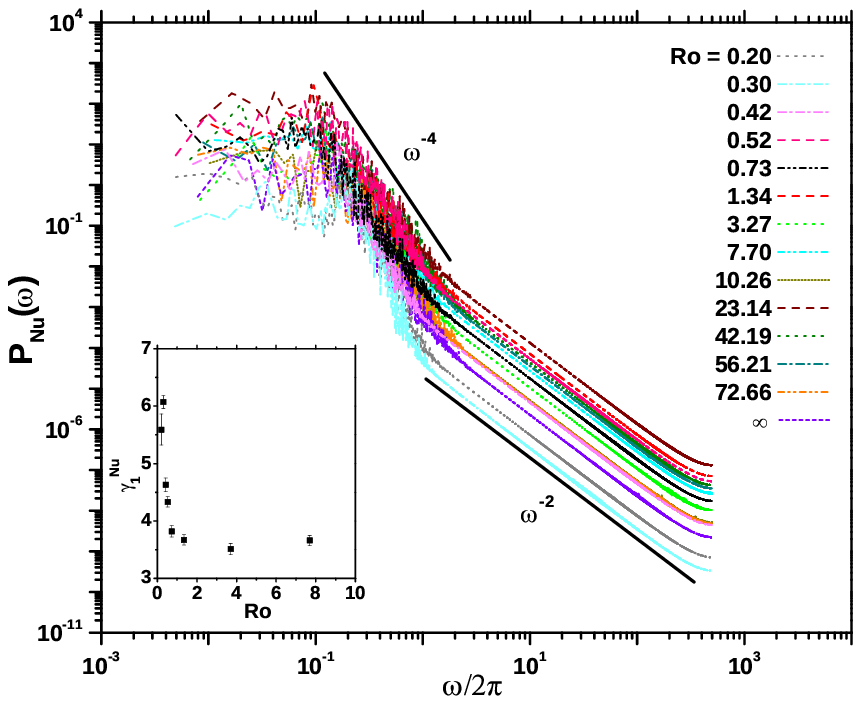}
\caption{(Color online) Power spectra of temporal fluctuations of the heat flux (Nusselt number $\mathrm{Nu}$), as computed from DNS, for different values  of $\mathrm{Ro}$. The power spectrum $P_{\mathrm{Nu}} (\omega)$ shows two scaling regimes: $P_{\mathrm{Nu}} (\omega)$ scales as $\omega^{-\gamma_1^{\mathrm{Nu}}}$ with $3.57 \le \gamma_1^{\mathrm{Nu}} \le 6.35$ at intermediate frequencies and $\omega^{-\gamma_2^{\mathrm{Nu}}}$ with $\gamma_2^{\mathrm{Nu}} \sim 2$ at higher frequencies.  The inset shows the variation of $\gamma_1^{\mathrm{Nu}}$ with $\mathrm{Ro}$.} \label{nu_spectra} 
\end{figure*}

We present the results of direct numerical simulations (DNS) on the frequency spectra of the convective entropy $E_{\Phi} (\omega)$ and the kinetic energy $E (\omega)$ along with the power spectrum of the heat flux $P_{\mathrm{Nu}} (\omega)$ for low-Prandtl-number Rayleigh-B\'{e}nard convection ($\mathrm{Pr} < 0.6$) with uniform rotation about the vertical axis.  Simulations are done for a wide range of the convective Rossby number $\mathrm{Ro} = \sqrt{\mathrm{Ra/(Ta Pr)}}$ ($0.2 \le \mathrm{Ro} < \infty$). The power spectra of all the three quantities show two scaling regimes: The spectra scale with frequency $\omega$ as $\omega^{-\gamma_1}$ at intermediate frequencies and as $\omega^{-2}$ at higher frequencies. The scaling exponent $\gamma_1$ depends on $\mathrm{Ro}$, $\mathrm{Ra}$ and $\mathrm{Pr}$. It is found that $3.0 < \gamma_1 < 4.4$ for $\mathrm{Ro} > 1$ and  $4.0 < \gamma_1 < 6.6$ for $0.2 \le \mathrm{Ro} < 1$.  There is no scaling behavior observed at lower frequencies.

\begin{figure}[ht]
\includegraphics[height=6.5 cm, width=8.0 cm]{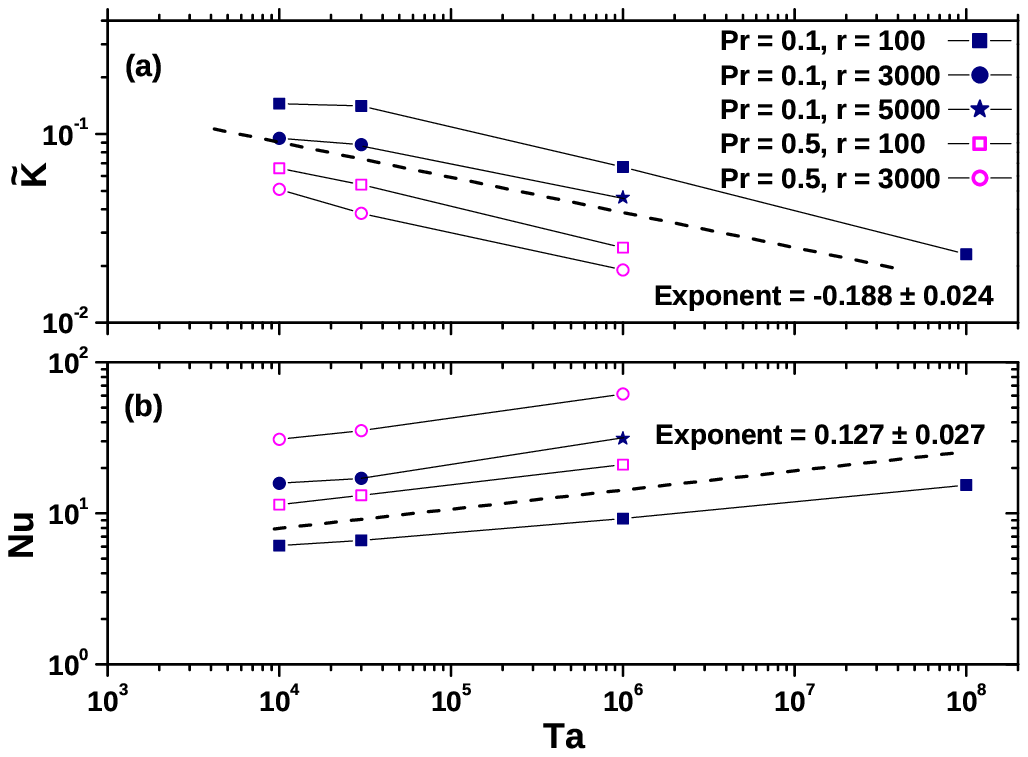}
\caption{(Color online) (a) Mean energy $\tilde{K}$ and (b) Nusselt number $\mathrm{Nu}$ scale with Taylor number $\mathrm{Ta}$ as ${\mathrm{Ta}}^{-0.19}$ and ${\mathrm{Ta}}^{0.13}$, respectively.} \label{k_nu_ta} 
\end{figure}

We consider thermal convection in a thin horizontal layer of fluid of thickness $d$, kinematic viscosity $\nu$, thermal diffusivity $\kappa$, and rotating uniformly about a vertical axis with angular velocity $\Omega$. A temperature difference of $\Delta T$ is imposed across the fluid layer. The convective dynamics is controlled by three dimensionless numbers that control the convective flow dynamics: the Rayleigh number $\mathrm{Ra} = \alpha_0 {\Delta T} g d^3/(\nu \kappa$), the Taylor number $\mathrm{Ta} = 4\Omega^2  d^4/\nu^2$, and the Prandtl number $\mathrm{Pr} = {\nu/\kappa}$. We consider thermally conducting and {\it stress-free} conditions on the top and bottom boundaries. This leads to the boundary conditions: $\partial_{z}v_1 = \partial_{z}v_2 = v_3 = \theta =0$ at $z=0$ and $z=d$. The boundary conditions in the horizontal direction are considered to be periodic. We then numerically integrate the dimensionless hydrodynamic system using pseudo spectral method~(see for details ~\cite{pharasi_etal_2014}). We use long signals for more than $150$ dimensionless time units with grid resolutions of $256^3$ for this purpose. The data points from the computed signals are recorded at equal time steps of $0.001$ to determine frequency spectra. A time-dependent convection appears at onset as $\mathrm{Ra}$ is raised above a critical value $\mathrm{Ra}_{\circ} (\mathrm{Ta}, \mathrm{Pr})$ for the values of $\mathrm{Ta}$ and $\mathrm{Pr}$ considered here. The flow becomes unsteady even at smaller values of the reduced Rayleigh number $r = \mathrm{Ra}/\mathrm{Ra}_{\circ} (\mathrm{Ta}, \mathrm{Pr}) > 5$ for $\mathrm{Pr} < 0.6$.  

Figure~\ref{entropy} displays the frequency spectrum of the entropy of the turbulent flow $E_{\Phi} (\omega)$, which scales with $\omega$ approximately as $\omega^{-\gamma_1^{\Phi}}$ with $3.4 \le \gamma_1^{\Phi} \le 6.6$ at intermediate values of the dimensionless frequencies ($0.07 < \omega/(2\pi) < 0.70 $). The scaling exponent $\gamma_1^{\Phi}$ shows a relatively weak dependence on $\mathrm{Ro}$, $\mathrm{Pr}$, $\mathrm{Ta}$ and $r$ at intermediate frequencies for $\mathrm{Ro} > 1$. The value of $\gamma_1^{\Phi}$ obtained here is in good agreement with the scaling exponent observed in an experiment without rotation by Cioni {\it et al.}~\cite{cioni_etal_pre_1997} for turbulent convection in a very low-Prandtl-number fluid. The exponent $\gamma_1^{\Phi}$ increases sharply as $\mathrm{Ro}$ decreases (see the inset in Fig.~\ref{entropy}) for $\mathrm{Ro} < 1$. The exponent $\gamma_1^{\Phi}$ varies between $4$ and $6.6$ for $0.2 \le \mathrm{Ro} < 1$. The unsteady convection at smaller values of $\mathrm{Ro}$ (typically, $r < 50$) is unlikely to be in a statistical steady state. This makes the exponent $\gamma_1^{\Phi}$ non-universal for flows at $\mathrm{Ro} < 1$.  The spectrum shows a different scaling at higher frequencies ($0.7 \le \omega/(2\pi) \le 200$). $E_{\Phi} (\omega)$ clearly scales with $\omega$ as $\omega^{-\gamma_2^{\Phi}}$ with $1.96 \le \gamma_2^{\Phi} \le 1.98$. The power spectra of the temperature fluctuations showing $\omega^{-2}$ behavior was clearly observed in experiments by Boubnov and Golitsyn ~\cite{boubnov_golitsyn}. The scaling extends for almost two decades, and it is independent of $\mathrm{Ro}$, $\mathrm{Pr}$, $\mathrm{Ta}$ and $r$ at higher frequencies. Zhou and Xia~\cite{zhou_xia_prl_2001} [see Fig.~1 (b) therein] observed $E_{\Phi} (\omega) \sim \omega^{-2}$  at higher values of $\mathrm{Ra}$ in the absence of rotation. The exponents $\gamma_1^{\Phi}$ and $\gamma_2^{\Phi}$ extracted by fitting the data for entropy spectra obtained from DNS are listed in Table~\ref{table1} for several values of $\mathrm{Ro}$, $\mathrm{Pr}$, and $r$. 

Figure~\ref{energy} displays the frequency spectrum  of the  energy $E(\omega)$ for different values of $\mathrm{Ro}$. The $E(\omega)$ also shows two scaling regimes. It scales with $\omega$ as $\omega^{-\gamma_1}$ with $3.06 \le \gamma_1 \le 6.54$ at intermediate frequencies. The scaling exponent $\gamma_1$ shows a weak dependence on $\mathrm{Ro}$, $\mathrm{Pr}$, $\mathrm{Ta}$ and $r$ for $\mathrm{Ro} > 1$.  Like $\gamma_1^{\Phi}$, $\gamma_1$ increases sharply with a decrease in $\mathrm{Ro}$ for $\mathrm{Ro} < 1$ (Fig.~\ref{energy}, inset). The energy power spectrum shows very clear scaling at higher frequencies. $E(\omega)$ scales with $\omega$ as $\omega^{-\gamma_2}$ with $\gamma_2 \sim 2$. The scaling exponent $\gamma_2$, which is independent of $\mathrm{Ro}$, $\mathrm{Pr}$, $\mathrm{Ta}$, and $r$, extends for more than two decades in $\omega$. The results of the best fit for $\gamma_1$ and $\gamma_2$ are listed in the seventh and eighth columns of Table~\ref{table1}. A comparison of crossover frequencies shows that $\omega_K < \omega_{\Phi}$, as had been anticipated for $\mathrm{Ro} \rightarrow \infty$. There is a weak dependence of $c_0$ on $\mathrm{Ro}$. 

We have also computed the power spectrum of the heat flux ($\mathrm{Nu}$) across the fluid layer. The power spectrum $P_{\mathrm{Nu}} (\omega)$ also shows two scaling regimes (Fig.~\ref{nu_spectra}). It scales with $\omega$ as $\omega^{-\gamma_1^{\mathrm{Nu}}}$ with $3.57 < \gamma_1^{\mathrm{Nu}} < 6.35$ at intermediate frequencies, and it scales  as $\omega^{-\gamma_2^{\mathrm{Nu}}}$ with $\gamma_2^{\mathrm{Nu}} \sim 2$ at high frequencies. The experiments by Auma\^{i}tre and Fauve~\cite{aumaitre_fauve_epl_2003} showed that the power spectra of the temporal fluctuations of the heat flux scaled as $\sim \omega^{-2}$ in the central part of the Rayleigh-B\'{e}nard cell and as $\sim \omega^{-4}$ near the boundaries in the absence of rotation. The results of the best fit are listed in the ninth and the tenth columns of Table~\ref{table1}. The exponent $\gamma_1^{\mathrm{Nu}}$ increases as $\mathrm{Ro}$ decreases for $\mathrm{Ro} < 1$ (Fig.~\ref{nu_spectra}, inset), as in the case of frequency spectra of the entropy and energy. The flow is far away from fully developed turbulence. We do not have an answer based on dimensional analysis for scaling exponents of the power spectrum of $\mathrm{Nu}$.

Variations of the time averaged energy $\tilde{K}$ and $\mathrm{Nu}$  with $\mathrm{Ta}$ are shown in Fig.~\ref{k_nu_ta} for different values of $\mathrm{Pr}$ and $r$. The time averaged energy $\tilde{K}$ varies with $\mathrm{Ta} \sim \mathrm{Ta}^{-0.2}$ for fixed values of $\mathrm{Pr}$ and $r$. It is consistent with the scaling used for the simulations. The dimensionless  energy $\tilde{K}$ $=$ $K d^2/(\nu \kappa \mathrm{Ra})$ $=$ $K d^2/[\nu \kappa r \mathrm{Ra_c (Ta)}]$ decreases with $\mathrm{Ta}$ as $\mathrm{Ra_c (Ta)}$ increases with $\mathrm{Ta}$ for a fixed values of $r$. The Nusselt number, on the other hand, varies with $\mathrm{Ta}$ as $\mathrm{Ta}^{0.13}$ for fixed values of $r$ and $\mathrm{Pr}$. This is also consistent, as $\mathrm{Ra}$ has to be increased with increasing $\mathrm{Ta}$  at a fixed value of $r$. This leads to an increase in $\mathrm{Nu}$ with $\mathrm{Ta}$. 

We have presented the scaling properties of the frequency spectra of convective entropy and kinetic energy along with the power spectrum of the heat flux for the turbulent flow in low-Prandtl-number Rayleigh-B\'{e}nard convection with uniform rotation. The spectra show two very different scaling regimes: $\omega^{-\gamma_1}$ at moderate frequencies and $\omega^{-\gamma_2}$ at relatively higher frequencies. The scaling exponent $\gamma_1$ at moderate frequencies shows a weak dependence on $\mathrm{Ro}$ for $\mathrm{Ro}>1$ with $\gamma_1 \sim 4$,  but shows a strong dependence on $\mathrm{Ro}$ for $\mathrm{Ro}<1$. The scaling exponent $\gamma_2 \approx 2$ at higher frequencies is independent of $\mathrm{Ro}$.  The frequency spectra of global quantities rather than local ones appear to make both the Kolmogorov and Bolgiano-like regimes more accessible. 

\noindent We thank S. Fauve (ENS, Paris) for fruitful discussions.


\begin{thebibliography}{100}
\bibitem{K41}
{A.N. Kolmogorov}, {Dokl. Akad. Nauk. SSSR} {\bf 30}, {299} {(1941)}; reprinted in {Proc. R. Soc. Lond. A} {\bf 434}, 15 (1991).
\bibitem{turbulence_review}
{E.D. Siggia}, {Annu. Rev. Fluid Mech.} {\bf 26}, {137} {(1994)}; {G. Ahlers, S. Grossmann, and D. Lohse}, {Rev. Mod. Phys.} {\bf81}, {503} {(2009)}.
\bibitem{lohse_etal_review}
{D. Lohse and K.-Q. Xia}, {Annu. Rev. Fluid Mech.} {\bf 42}, {335} {(2010)}.
\bibitem{vsl_1991}
{V.S. L'vov}, {Phys. Rev. Lett.} {\bf 67}, {687} {(1991)}.
\bibitem{BO}
{R. Bolgiano}, {J. Geophys. Res.} {\bf 64}, {2226} {(1959)}; {A.M. Obukhov}, {Dokl. Akad. Nauk. SSSR} {\bf 125}, {1246} {(1959)}.
\bibitem{pharasi_etal_2014}
{H.K. Pharasi, K. Kumar, and J.K. Bhattacharjee}, {Phys. Rev. E} {\bf 89}, {023009} {(2014)}, and references therein.
\bibitem{cioni_etal_pre_1997}
{S. Cioni, S. Horanyi, L. Krebs, and U. M\"{u}ller}, {Phys. Rev. E} {\bf 56}, {R3753} {(1997)}.
\bibitem{boubnov_golitsyn}
{B. M. Boubnov and G. S. Golitsyn}, {J. Fluid Mech.} {\bf 219}, {215} {(1990)}.
\bibitem{zhou_xia_prl_2001}
{S.-Q. Zhou  and K.-Q. Xia }, {Phys. Rev. Lett.} { \bf 87}, {064501} {(2001)}.
\bibitem{aumaitre_fauve_epl_2003}
{S. Auma\^{i}tre and S. Fauve}, {Europhys. Lett.} {\bf 62}, {822} {(2003)}.

\end{thebibliography}
\end{document}